\documentclass[runningheads]{llncs}
\usepackage[T1]{fontenc}
\usepackage{graphicx}
\usepackage{xcolor}
\usepackage{subcaption}
\usepackage{listings}
\usepackage{amsmath}
\usepackage{amsfonts}
\usepackage{semantic}
\usepackage{hyperref}

\usepackage[pscoord]{eso-pic}
\newcommand{\placetextbox}[4]{
  \setbox0=\hbox{#4}
  \AddToShipoutPictureFG*{
    \if#3r
    \put(\LenToUnit{\paperwidth-#1},\LenToUnit{\paperheight-#2}){\vtop{{\null}\makebox[0pt][r]{\begin{tabular}{r}#4\end{tabular}}}}%
    \else
    \put(\LenToUnit{#1},\LenToUnit{\paperheight-#2}){\vtop{{\null}\makebox[0pt][l]{\begin{tabular}{l}#4\end{tabular}}}}%
    \fi
  }%
}%

\lstdefinestyle{mystyle}{
    breakatwhitespace=false,         
    basicstyle=\ttfamily\small,
    breaklines=true,                 
    columns=fullflexible,
    captionpos=b,                    
    keepspaces=true,                 
    showspaces=false,                
    showstringspaces=false,
    showtabs=false,                  
}

\lstdefinelanguage{certus}
{
    keywords=[1]{
        if,
        else,
        then,
        cases,
        otherwise
    },
    keywordstyle=[1]\color{blue},
    keywords=[2]{
        and,
        or,
        \&,
        |
    },
    keywordstyle=[2]\color{black},
    keywords=[3]{
        is,
        overlaps,
        contains,
        gt,
        lt
    },
    keywordstyle=[3]\color{violet},
    keywords=[4]{
        intersect,
        union,
        majority,
        minority,
        lowest,
        highest,
        not,
        up,
        down,
        trap,
        crisp
    },
    keywordstyle=[4]\color{teal},
    keywords=[5]{
        with,
        global,
        as,
        because
    },
    keywordstyle=[5]\color{orange},
    sensitive=false,
    morestring=[b]',
    morecomment=[l]{--}
}

\newcommand{\bb}[1]{\textbf{#1}}
\renewcommand{\it}[1]{\textit{#1}}
\renewcommand{\tt}[1]{\texttt{#1}}

\newcommand{\ul}[1]{\underline{#1}}

\newcommand{\bel}{\alpha}
\newcommand{\C}{\mathcal{C}}

\newcommand{\Certus}{\it{Certus}}

\newcommand{\il}[1]{\lstinline{#1}}

\newcommand{\certain}{\it{certain}}
\newcommand{\vhigh}{\it{very high}}
\newcommand{\high}{\it{high}}
\newcommand{\med}{\it{med}}
\newcommand{\low}{\it{low}}
\newcommand{\vlow}{\it{very low}}
\newcommand{\zero}{\it{zero}}

\newcommand{\squeezeHeader}{\vspace{-0.5em}}

\begin{document}
\title{Certus: A domain specific language for confidence assessment in assurance cases}
\titlerunning{Certus: A domain specific language for confidence assessment}
%
\author{Simon Diemert\inst{1,2}\orcidID{0000-0001-9493-7969} \and
Jens H. Weber\inst{1}\orcidID{0000-0003-4591-6728}}
\authorrunning{S. Diemert and J. H. Weber}
%
\institute{University of Victoria, Victoria, Canada
\and Critical Systems Labs Inc., Vancouver, Canada}

\maketitle 
\begin{abstract}
Assurance cases (ACs) are prepared to argue that a system has satisfied critical quality attributes. Many methods exist to assess confidence in ACs, including quantitative methods that represent confidence numerically. While quantitative methods are attractive in principle, existing methods suffer from issues related to interpretation, subjectivity, scalability, dialectic reasoning, and trustworthiness, which have limited their adoption. This paper introduces \Certus{}, a domain specific language for quantitative confidence assessment. In \Certus{}, users describe their confidence with fuzzy sets, which allow them to represent their judgment using vague, but linguistically meaningful terminology. \Certus{} includes syntax to specify confidence propagation using expressions that can be easily inspected by users. To demonstrate the concept of the language, \Certus{} is applied to a worked example from the automotive domain. 

\keywords{Assurance Cases \and Safety Cases \and Confidence Assessment \and Domain Specific Language \and Fuzzy Sets}
\end{abstract}
\placetextbox{8cm}{24cm}{r}{{\color{red}\bb{Preprint. Submitted to SASSUR'25}}}
\section{Introduction}\label{sec:intro}

During the development of a critical system, engineers often prepare an Assurance Case (AC), as a ``reasoned and compelling argument, supported by a body of evidence, that a system, service or organisation will operate as intended for a defined application in a defined environment'' \cite{gsn}. ACs are usually focused on a single quality attribute, such as safety (i.e., a ``safety case'') or security (i.e., ``security-case''). Preparing an AC is required for compliance with a range of industrial standards; a notable and recent example being ISO/PAS 8800 which identifies preparing an AC as a central pillar in the AI assurance process for automotive technology \cite{iso8800}. Several notations exist for organizing the AC's argument, including the Goal Structuring Notation (GSN), Claims-Argument-Evidence (CAE), Eliminative Argumentation (EA), and the Friendly Argument Notation (FAN) \cite{gsn,kelly,goodenough2015,cae}. While notation provides a foundation for describing AC arguments in terms of syntax and basic semantics, this is not enough to determine if an AC is acceptable. A question arises: \it{is there sufficient confidence that the claims in the AC are true?}

Many Confidence Assessment Methods (CAMs) for ACs exist. Some CAMs are qualitative \cite{hawkins2011,fenn2024,goodenough2015,holloway2021} whereas others are quantitative and produce a numerical valuation of confidence in the claims in the argument \cite{hobbs2012,idmessaoud2024,herd2024}, and some mix qualitative and quantitative aspects \cite{bloomfield2023}. Quantitative CAMs are attractive because they ``sum up'' the confidence in a claim into a number (or handful of numbers), which has benefits in terms of communication with interestholders and higher-level (potentially automated) decision-making about critical systems. However, there are also concerns about quantitative CAMS arising from  the methods themselves and their validation \cite{diemert2024cams,graydon2017}.

This paper introduces \Certus{}, a domain specific language (DSL) for AC confidence assessment that uses fuzzy sets to represent confidence in the argument and its supporting evidence. Using fuzzy sets, authors of ACs can express their confidence using vague, but linguistically meaningful, terms like ``high'' or ``very low'' confidence. Additionally, \Certus{} authors can express propagation operations that compute confidence in a parent node based on its children. In prior work we suggested expressing confidence linguistically but did not formalize the concept \cite{diemert2023}. To our knowledge, \Certus{} is the first proposal to both model confidence in an AC with fuzzy sets and propagate confidence with a DSL.

The remainder of this paper is structured as follows. First, Section \ref{sec:motivation} discusses limitations of existing quantitative CAMs. Next, Section \ref{sec:language} describes our approach for modeling confidence using fuzzy set and presents the \Certus{} language. Section \ref{sec:example} applies \Certus{} to small fragment from a larger automotive AC. Finally, Section \ref{sec:discussion} closes with an outline for future work and concluding remarks.

Before proceeding, we note that this paper does not aim to compare qualitative or quantitative CAMs, nor does it take a position on whether qualitative or quantitative CAMs are preferred. In fact, as with many areas of engineering, different tools are applicable in different contexts and the choice to use a specific CAM depends on many factors. Our aim in this paper is to propose a tool to address known weaknesses of quantitative CAMs. 
\section{Motivation}\label{sec:motivation}

Despite many quantitative CAMs existing in the literature, they appear to have limited use in practice. Possible reasons for a lack of adoption are introduced below, based on previous work and our own practical experiences \cite{graydon2017,diemert2024cams}. Addressing these challenges is the motivation for the \Certus{} project.

\paragraph{Interpretation.} Quantitative CAMs produce a numerical valuation of confidence that must be interpreted by decision makers. There are at least two challenges with interpretation. First, it can be difficult to determine whether the calculated level of confidence is acceptable (e.g., is 0.93 confidence acceptable? Why not 0.92 or 0.94?). Second, numbers can easily be taken out of context or misinterpreted by non-experts (e.g., ``But 0.93 confidence means there is 0.07 probability of system failure!''). Ideally, quantitative CAMs should represent results in a manner that resists misinterpretation or have interpretation guides readily available.

\paragraph{Subjectivity.} Many existing quantitative CAMs require users to express judgement precisely, as one or more numbers. This is a subjective activity due to the natural variability in human judgement. Even experts in the same field might ascribe different weights to evidence or arguments based on their education and experience. We hypothesize that reducing confidence to numbers and ``one size fits all'' calculations result in users loosing (qualitative) nuance or conceptual depth that is often important in matters of engineering judgement. Ideally, quantitative CAMs should allow judgement to be expressed vaguely and in a manner that is flexible enough to capture nuanced reasoning.

\paragraph{Scaling.} Applying quantitative CAMs requires additional effort. For instance, to apply the Dempster-Shafer Theory method, a user must input four values per argument leaf node and at least three values per argument step \cite{idmessaoud2024}; when scaled to a large AC, this requires significant effort. Ideally, quantitative CAMs should either limit the number of inputs required, or provide mechanisms to reduce the number of inputs required in the most frequent use cases.

\paragraph{Defeaters.} Dialectic reasoning (aka ``defeaters'') are increasingly used by practitioners to reason about doubt in an AC. However, despite the role of defeaters in expressing reasons to reduce confidence, few methods accommodate defeaters \cite{diemert2024,herd2025}. Ideally, quantitative CAMs should be capable of handling the ``negative confidence'' expressed by defeaters.

\paragraph{Trustworthiness.} There are at least two issues relating to trustworthiness for quantitative CAMs. First, the mathematic frameworks used are complex, making them less accessible to busy practitioners. Tools that help automate calculations become ``black boxes''. As a result, practitioners are less likely to place their trust in a method where the means of calculation are not easily understood. Second, there is a lack of empirical evidence (e.g., case studies and controlled experiments) showing that methods produce trustworthy results \cite{graydon2017}. Ideally, quantitative CAMs should be based on easily understood mathematical theories (or otherwise provide accessible guidance), and be validated to demonstrate they produce repeatable results that match intuition.

\paragraph{Summary.} In summary, there are several limitations or challenges that prevent use of quantitative CAMs. Addressing them, in whole or in part, is our motivation for creating \Certus{}. Fuzzy sets are a strong candidate for addressing challenges related to interpretation and subjectivity due to their ability to represent vague, but linguistically meaningful, information \cite{zadeh}. Additionally, using a DSL for specifying confidence propagation will support users to describe more complex propagation rules, accommodate defeaters, scale to large ACs, and promote trustworthiness by increasing transparency for how confidence is propagated.

\section{The \Certus{} Language}\label{sec:language}

This section introduces the core concepts of the \Certus{} language, beginning with how to model AC confidence using fuzzy sets and then describing how to specify the propagation of confidence through an AC argument. For both brevity and clarity, we sketch the semantics of the language, and defer a formalization to future work. \Certus{} assumes that AC arguments can be modelled as Directed Acyclic Graphs (DAGs). We use a subset of the EA notation in our examples: claims are shown as rectangles, evidence appears in blue ovals, and defeaters are red irregular octagons \cite{goodenough2015}. We selected EA as a matter of preference; \Certus{} is applicable to other notations that represent arguments as a DAG.

\subsection{Modelling Confidence with Fuzzy Sets}

Let $\bel_i \in \beta$ represent the degree of confidence that claim $C_i$ is true, for any ordered set $\beta$ describing degrees of confidence. Usually we take $\beta = [0,1]$ with $1.0$ corresponding to maximum confidence (i.e., $\bel_i = 1.0$ reflects absolute confidence that $C_i$ is true) and $0.0$ is the minimum confidence (i.e., $\bel_i = 0.0$ means that there is no confidence in the truth of $C_i$).

\begin{figure}[t]
    \centering
    \begin{subfigure}{0.48\textwidth}
        \includegraphics[width=\textwidth]{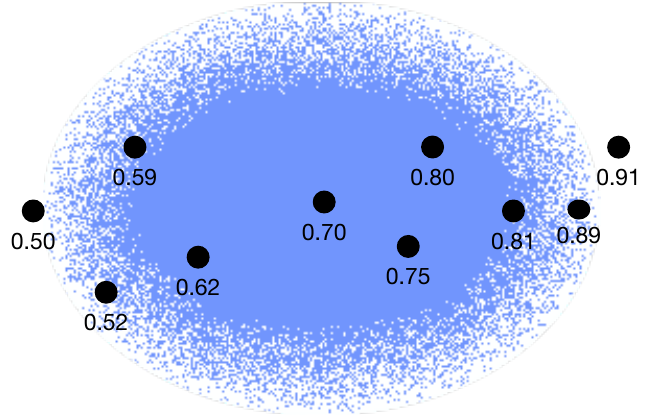}
    \end{subfigure}
    \hfill
    \begin{subfigure}{0.48\textwidth}
        \includegraphics[width=\textwidth]{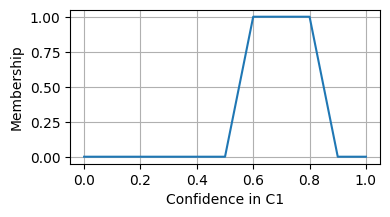}
    \end{subfigure}
    \caption{Visualization of a fuzzy set for ``'high'' confidence and plot of the corresponding fuzzy membership function.}
    \label{fig:fuzzyset}
\end{figure}

Consider the statement ``\it{my confidence in claim $C_1$ is \ul{high}}.'' This is a vague linguistic expression of confidence: how should it be interpreted? Specifically, what degree(s) of confidence in $\beta = [0,1]$ are considered ``high''? Suppose we decide degrees of confidence in $[0.6,0.8]$ to definitively correspond to high belief, and then degrees of belief in $[0.5,0.6)$ and $(0.8, 0.9]$ somewhat less corresponding to high belief. The left-hand diagram in Fig.~\ref{fig:fuzzyset} visualizes this scenario. Degrees of confidence are annotated as points: the points (e.g., 0.70, 0.62, 0.75) in the ``core'' are definitely in the set; some points are only partially in the set (e.g., 0.59, 0.52); and some points are entirely outside the set (e.g., 0.91 and 0.50). The same information can be shown as a membership function on the degrees of confidence in claim $C_1$, which appears on the right-hand side of Figure \ref{fig:fuzzyset}. Denote this fuzzy set membership function for ``high'' belief as $\mu_{high} : \beta \to [0,1]$

This membership function $\mu_{high}$ maps degrees of confidence in a claim into membership in a fuzzy set denoting a ``high'' level of confidence. In doing so, it gives meaning to the statement: \it{confidence in $C_1$ is \ul{high}}. In other words, the fuzzy set $\mu_{high}$ characterizes one's confidence in the truth of the claim $C_1$. This formulation can be generalized to arbitrary fuzzy sets for describing confidence in a claim, allowing for statements like: \it{confidence in $C_1$ is \ul{A}}, for some vague qualifier that can be encoded as a fuzzy set membership function. Throughout this paper, we denote the set of all confidence describing fuzzy sets on the domain $\beta$ (usually $[0,1]$) as $\C$.

\begin{figure}
    \centering
    \includegraphics[width=0.8\textwidth]{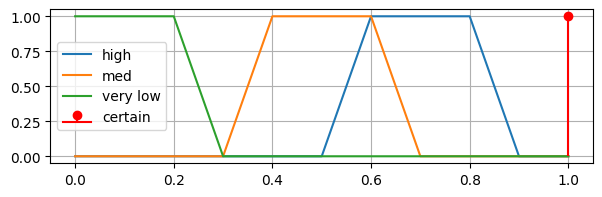}
    \caption{Some canonical sets defined by the \Certus{} language.}
    \label{fig:canonical}
\end{figure}

In \Certus{} users can define any convex and normalized fuzzy set to describe their confidence. However, it is convenient to have a set of pre-defined canonical fuzzy sets that represent common confidence expressions. For the purpose of this paper, we identify five such sets: \zero{}, \vlow{}, \low{}, \med{}, \high{}, \vhigh{}, and \certain{}. For reference, some of these are depicted in Fig. \ref{fig:canonical}. Note that \zero{} and \certain{} are ``crisp'' singleton sets with only one member: $\{0\}$ and $\{1\}$, respectively. 

\subsection{Propagating Confidence with \Certus{}}

\begin{figure}
    \centering
    \includegraphics[width=0.9\textwidth]{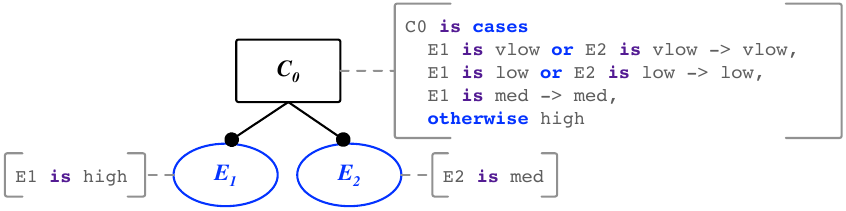}
    \caption{Simple propagation step in \Certus{}.}
    \label{fig:prop-simple}
\end{figure}

Building on the ability to describe confidence in a claim using fuzzy sets, the two fundamental aspects of the \Certus{} language can be defined: confidence assignment and confidence propagation. Confidence is assessed upward through the DAG in a recursive manner, from confidence assignment (usually the leaves) to the top-level node. At each logical step in the argument, a propagation operator is used to determine the confidence in the parent node based on the confidence assessed for the children. An example for a single argument step is shown in Fig.~\ref{fig:prop-simple}, where \Certus{} annotations appear within partial rectangles connected to nodes with a dashed line. In this example, leaf nodes $E_1$ and $E_2$ have their confidence assigned and the parent node, $C_0$, has a confidence propagation operator. When assessed by \Certus{}, the confidence in $C_0$ will be \high{}.

\squeezeHeader{}
\paragraph{Confidence Assignment.} Confidence assignments allow a user to specify their confidence in an argument's nodes using a simple expression (e.g., \il{E1 is high}). The result is that the canonical fuzzy set \high{} is associated with the evidence node $E_1$. For \Certus{} to assess confidence in an AC, every path in the argument's DAG must include a confidence assignment from which confidence assessment can begin. It is preferable to apply assignments at the leaves of the DAG, which usually correspond to evidence typed nodes (i.e., ``\it{how confident are we in this piece of evidence?}''). However, it is also possible to specify confidence on a non-leaf node, which we call ``shorting''\footnote{This is a reference to short-circuiting an electrical circuit and also short-circuiting logical operators in some programming languages.}. In this case, \Certus{}'s confidence assessment will not proceed further down the path and the shorting assignment will be used for propagation instead. Shorting is to be discouraged, because it ignores the arguments below the shorted node and might result in confidence assessments that are disconnected from real-world observations about a system. Even so, it can be useful in practice, especially assessing large or complex arguments.


\squeezeHeader{}
\paragraph{Confidence Propagation.} Confidence propagation is an operation that, when applied to a single argument step, determines the confidence in a parent node based on the confidence(s) in its children. Propagation is performed either by: 1) direct child to parent assignment, or 2) using the \il{cases} expression. For direct propagation, the confidence of a single child node is assigned directly to the parent. This is denoted as \il{C0 is C1}.

The \il{cases} expression matches (in order of appearance) conditions on the child confidences. Logical connectors such as \il{and} and \il{or} are used to match multiple nodes' confidences in one condition in the usual manner. The right-hand side of a single case can either be a fuzzy set (e.g., \il{E1 is med -> med}) or the identifier of a node (e.g., \il{E2 <= low -> E1}). In the later case, the fuzzy set for the identified node is propagated upward.

The \il{cases} expression is the basis of all propagation operators in \Certus{}. This ensures that the core semantics are simple and easy for users to understand. Methods for specifying more complex and re-usable propagation operators are described below. The scope of a \il{cases} expression is limited to the direct descendants of the node it is applied to in the argument's DAG. We require that \il{cases} expressions in \Certus{} be total functions that map from the set of all in-scope confidence assignments to the set $\C$.


Other quantitative CAMs encode the relationship between a child and parents using numerical weight parameters that are an input to the confidence propagation formula(s) \cite{hobbs2012,idmessaoud2024}. This allows users to separate confidence in a premise (e.g., ``\it{I am confident this is high-quality evidence}'') from the strength of association between its parent claim (e.g., ``\it{I am confident this evidence supports the parent}''). In \Certus{}, there are no explicit weighting parameters. Instead, the user captures the relationship between a child and parent using the \il{cases} expression.

\subsection{Comparison Semantics in \Certus{}}

There are several comparison operators that can be used in \Certus{}'s conditional expressions to compare two fuzzy set membership functions. These are described below as binary relations on $\C$.

The \il{is} operator has two meanings in \Certus{}: assignment (described above) and comparison. Given two fuzzy sets $A$ and $B$, the comparison \il{A is B} evaluates to true if $A$ is a subset of (or equal to) $B$. More formally, $is : \C \times C \to \mathbb{B}$ such that $is(A,B)$ if $\forall x \in [0,1] : \mu_A(x) \leq \mu_B(x)$.

The \il{contains} operator is the reciprocal of \il{is}. Given two fuzzy sets $A$ and $B$, the comparison \il{A contains B} evaluates to true if $B$ is a subset of (or equal to) $B$. Formally, $contains : \C \times C \to \mathbb{B}$ such that $contains(A,B)$ if $\forall x \in [0,1] : \mu_A(x) \geq \mu_B(x)$.

The \il{overlaps} operator is much weaker than \il{is} and \il{contains}. Given two fuzzy sets $A$, and $B$, the comparison \il{A overlaps B} evaluates to true if they have some non-zero overlap in their membership functions. Formally, $overlaps : \C \times \C \to \mathbb{B}$ such that $overlaps(A,B)$ if $\exists x \in [0,1] : \mu_A(x) > 0 \land \mu_B(x) > 0$.

The \it{greater than} (denoted \il{>} or \il{gt}) and \it{less than} (denoted \il{<} or \il{lt}) operators require an ordering function to rank fuzzy sets. We use Yager's unit-interval fuzzy set ordering function to compare sets \cite{yager1981}. The ordering function, $F(A)$, computes the integral of the mean of level sets to produce a number that represents the position of the fuzzy set in $[0,1]$. Comparing these numbers for different fuzzy sets allows one to order the sets, i.e., declare one set to ``greater than'' another. The computation is: $F(A) = \int_0^1 M(A_{\alpha}) \; d\alpha$, where $A_\alpha = \{x : x \in [0,1], \mu_A(x) \geq \alpha \}$ is the $\alpha$-cut of the set $A$ (i.e., a ``level set'') and $M(A_\alpha)$ is the mean of the values in the level set. Then for fuzzy sets $A$ and $B$, the \it{greater than} operation is formally defined $gt : \C \times \C \to \mathbb{B}$ such that $gt(A,B)$ if $F(A) > F(B)$. And vice versa for the \it{less than} operation. These operators can be extended to check equality (e.g., \il{>=}, \il{<=}) by comparing the membership functions of the fuzzy sets directly. They also enable operations such as \il{min(...)} and \il{max(...)}, which have the usual definitions. It is worth noting that, while Yager's ordering function works well most of the time, it fails to correctly order sets in some scenarios where membership functions are non-convex or non-normal \cite{bortolan1985}.

\subsection{Defined Propagation Operators}

Even though the \il{cases} operator is flexible, it would be onerous to define lengthy propagation rules for every reasoning step in an AC. Therefore, \Certus{} allows users to define named propagation operators and invoke them by name. This is similar to the notion of a pre-defined function in many programming languages. There are two ways to define propagation operators in \Certus{}: parameterized operators and macro operators. Importantly, both use the \il{cases} expression described above.

\squeezeHeader{}
\paragraph*{Parameterized Propagation Operators.}

\begin{figure}[t]
    \centering
    \includegraphics[width=0.95\textwidth]{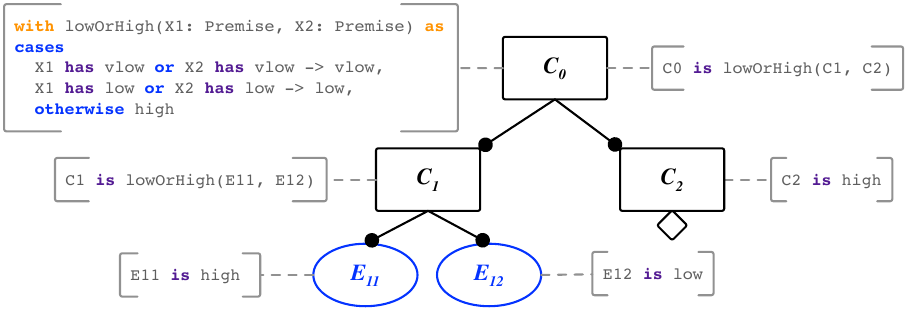}
    \caption{Using parameterized propagation operators.}
    \label{fig:paramOp}
\end{figure}

Using a parameterized operator, a user can define a propagation operator in advance and then invoke it by name. The operator must be defined at an ancestor node in the argument's DAG or as a global definition provided separately from the argument. The inputs to the operator are typed by argument node type. The user assigns the nodes to the parameters when the operator is used in an argument step. The syntax: \il{with <name>(<parameters>) as <definition>}, is used to define a parameterized propagation operator. This is demonstrated in Fig.~\ref{fig:paramOp}. An operator called \il{lowOrHigh} is defined to accept two nodes of type Premise (e.g., a Claim or Evidence nodes). If either of these nodes have fuzzy set membership functions that overlap the membership function of \low{} or \vlow{}, then the output of the operation is respectively \low{} or \vlow{}, otherwise the output is \high{}. The operator is used two times, once at node $C_1$ and again at node $C_0$. Using the assignments in the leaves of Fig.~\ref{fig:paramOp}, the overall confidence at $C_0$ is \low{}.

\squeezeHeader{}
\paragraph*{Macro Propagation Operators.}
The propagation operators described above depend on a fixed number of inputs (parameters or nodes in an argument step). As a result, they could be cumbersome to use in scenarios where the number or type of nodes are likely to change. To address this challenge, \Certus{} allows the user to define macros that are expanded into \il{cases} expressions as a pre-processing step before confidence is assessed. Macro expansion takes into account the current context of a node, including the number and type of children. Formally, macros are higher-order functions that map from the set of nodes in an argument step into the set of possible \il{cases} expressions. The expanded \il{cases} expressions can be inspected by users so that they can concretely understand how confidence is propagated through the argument.

\Certus{} does not currently have a macro definition language. Instead, it provides an interface to scripting languages, such as Python. To define a macro, the user must define a function that accepts a list of argument nodes and then returns a \il{cases} expression. The \il{\#MACRO\_NAME} syntax is used to invoke macros, which must match the name of a function defined in the scripting language. By convention, macro names are defined with uppercase letters.

\begin{figure}
    \centering
    \includegraphics[width=0.95\textwidth]{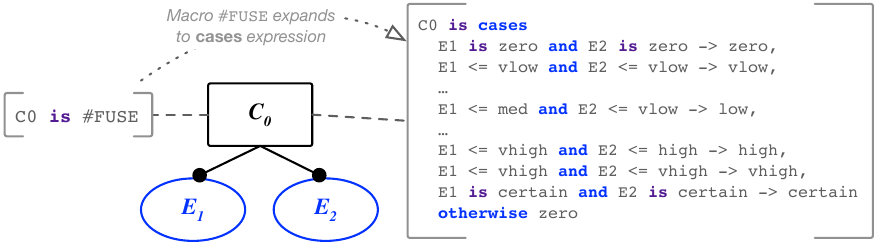}
    \caption{Example showing the expansion of the \il{FUSE} macro.}
    \label{fig:fuse-cases}
\end{figure}

At present, \Certus{} provides one built-in macro called \il{FUSE} that expands to a \il{cases} expression that merges (i.e., ``fuses'') multiple fuzzy sets together in a balanced manner. It uses an integer scoring function, $S : \{\zero{}, ..., \certain{}\} \to \mathbb{Z}$, to determine the output for each case in the expression, e.g., $S(\zero{}) = 0$ and $S(\certain{}) = 6$. Premise type nodes are assigned positive scores and defeater nodes are given negative scores. Then the average of all clauses in a given case is computed and used to determine the output set for that case using the same integer scale. The expansion of this \il{FUSE} is shown in Fig.~\ref{fig:fuse-cases} for the case of two nodes. For example, in the third case of the expanded expression we have: $\big(S(\med{}) + S(\vlow{})\big)/2 = \lfloor (3 + 1)/2 \rfloor =  S^{-1}(2) = \low{}$.

\subsection{Defeaters in \Certus{}}

As described above, dialectic elements (or ``defeaters'') are an important part of AC arguments, and quantitative CAMs should support defeaters in some manner. To include defeaters in \Certus{}, we consider both how to model them as fuzzy sets and how defeaters are to be handled as part of confidence propagation.

\squeezeHeader{}
\paragraph*{Modelling Defeaters with Fuzzy Sets.}
In the most basic sense, defeaters are negative premises that capture a reason to doubt their parent claim. Whereas a normal premise (e.g., claim or evidence) contributes positively to the confidence in a parent, a defeater decreases confidence in its parent. Like for positive premises, the confidence in the credibility of a defeater is also a matter of degree \cite{diemert2024}. It follows that confidence in a defeater can also be described as a fuzzy set. For instance, one could have \low{} confidence in a defeater, which corresponds to a scenario where a doubt is not very credible, but cannot be entirely ruled out. Conversely, a defeater with \vhigh{} confidence is likely (but not certainly) true.

\squeezeHeader{}
\paragraph*{Rules for Incorporating Defeaters.}
In prior work, we introduced 12 rules for incorporating defeaters into quantitative CAMs \cite{diemert2024}. CAMs like the BBN or DST methods use pre-determined formulas to propagate confidence through the argument. These formulas can be extended and then analyzed to demonstrate they handle defeaters appropriately. In \Certus{} the propagation operators are defined by the user and are not known in advance. Therefore, it is not possible to show \it{a priori} that confidence propagation in \Certus{} respects the 12 rules for defeaters. However, it is still desirable for defeaters to be managed as an integral part of the method. So, immediately prior to computing confidence, after all macros have been expanded, \Certus{} performs a static analysis of all \il{cases} expressions to confirm they comply with the 12 rules for defeaters. This applies to user-created expressions and expressions generated by expanding macros. These ``pre-flight checks'' are analogous to the checks performed by a compiler for many programming languages.

\section{Worked Example}\label{sec:example}

As a means of preliminary validation we have implemented a proof-of-concept version of \Certus{}\footnote{\url{https://gitlab.com/sdiemert/certus}} and applied it to an AC fragment from an exemplar automotive adaptive cruise control (ACC) system, which is shown below in Fig.~\ref{fig:example}. \Certus{} annotations appear in partial rectangles as above, and the name of the computed fuzzy for each node is provided for reference in the top-right of each node. The remainder of this section highlights points of interest in the example.

\begin{figure}[t]
    \centering
    \includegraphics[width=\textwidth]{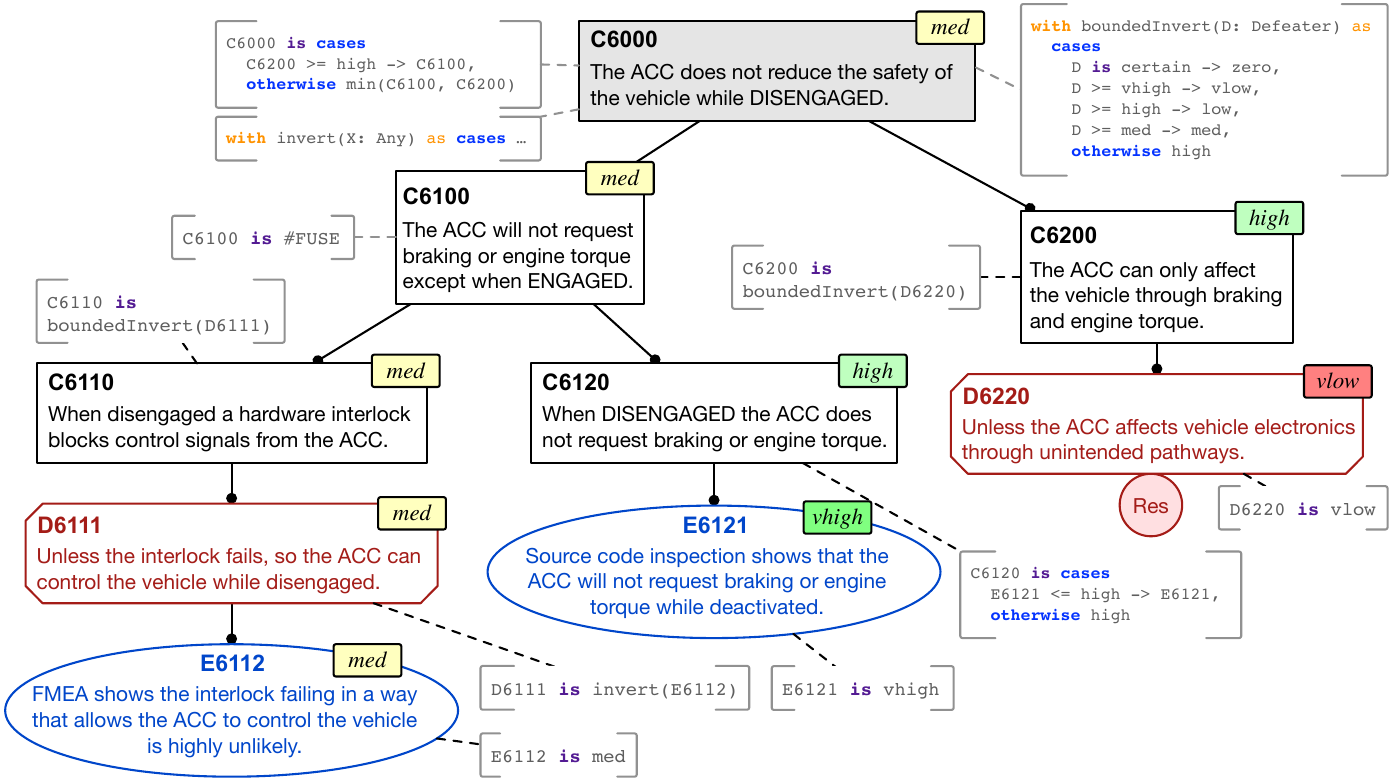}
    \caption{Fragment from ACC AC with \Certus{} applied.}
    \label{fig:example}
\end{figure}

First, the example shows the use of two parameterized propagation operators: \il{invert} and \il{boundedInvert}. The \il{invert} operation returns the inverted confidence of its input (e.g., \low{} becomes \high{}), it is not fully defined in Fig.~\ref{fig:example} for brevity. The \il{boundedInvert} is intended for reasoning steps where a premise is challenged by a single defeater. It is ``bounded'' in the sense that it does not output confidence greater than \high{}, even if the input confidence is \vlow{}. Bounding the confidence that a parent derives from a single child defeater reflects the intuition that, in the absence of additional evidence to support the parent, there is a limit to the confidence that can be gained from resolving defeaters. 

Second, the \il{cases} expression for node \il{C6120} was designed to limit credit that can be taken from \il{E6121} to at most \high{} confidence, regardless of the confidence in the evidence. In this case, the limitation is justified on the basis that source code inspection on its own would be inadequate to achieve \vhigh{} confidence in the correct behaviour of a software function.

Third, the \il{cases} expression for node \il{C6000} weights \il{C6100} more heavily than \il{C6200}. Provided that \il{C6200} has at least \high{} confidence, then the confidence in \il{C6000} depends on \il{C6100}, otherwise the minimum confidence from the two branches is used. Examining the text of nodes \il{C6100} and \il{C6200} justifies this logic. \il{C6200} can be defeated if there is a case where an unknown interaction of the ACC with the other vehicle systems exists. Provided such an interaction is ruled out with reasonable confidence, then the remainder of the argument rests on whether ACC is well-behaved within its known scope of operation, as contemplated by \il{C6100}.

Finally, to compare with the existing BBN method, we used the same AC fragment as in \cite{diemert2024}, and we mapped the leaf-level confidences from the BBN example to fuzzy sets for this example (e.g., belief in \il{E6112} was $0.6$, which became \med{}). The BBN example had an overall confidence of $0.39$, which is consistent with, but perhaps on the low side, of the result produced with \Certus{}.
\section{Discussion}\label{sec:discussion}

This paper has introduced \Certus{}, a DSL for specifying confidence propagation in ACs using fuzzy sets. \Certus{} addresses some of the challenges and limitations that prevent AC practitioners for using quantitative CAMs on real-world ACs. 

Modelling confidence using fuzzy sets addresses challenges related to interpretation and subjectivity by allowing users to describe confidence using vague, but linguistically meaningful, expressions. For instance, in \Certus{}, users do not need to select specific numbers to represent confidence; instead, they select a fuzzy set (e.g., \vhigh{}) to represent their confidence. The output of \Certus{} is similarly a fuzzy set which can be associated with a linguistic expression for interpretation by decision makers.

By using a DSL to specify confidence propagation in an AC, \Certus{} allows users to represent sophisticated propagation rules that capture their reasoning and that accommodate dialectic reasoning. Further, \Certus{}'s capability to define and re-use propagation operators reduces the number of inputs required and helps with scaling to larger ACs. Finally, in \Certus{} confidence propagation operations are either represented by, or can be reduced to, simple expressions that can be inspected and understood by users, prompting trustworthiness. 

This paper is our first description of \Certus{}. There is significant work required to develop the concept and DSL prototype into a practice-ready CAM. From the perspective of developing the language, we plan to create a formal specification for the language; explore graphical representations of confidence propagation operators to improve readability; and create additional built-in macros for common operations. From an evaluation perspective, studies are required to show that \Certus{} is usable for practitioners and produces results that are trustworthy such that it can be used to support decision-making for critical systems.



%
%
%
\bibliographystyle{splncs04}
\bibliography{refs.bib}
\end{document}